\documentclass[epj]{webofc}
\usepackage[varg]{txfonts}   
\wocname{EPJ Web of Conferences}
\woctitle{ICNFP 2015}
\begin{document}
\selectlanguage{english}
\title{Recent results from NA61/SHINE}
%
%

\author{Evgeny Andronov\inst{1}\fnsep\thanks{\email{evgeny.andronov@cern.ch}}, On behalf of the NA61/SHINE Collaboration 
}

\institute{Saint Petersburg State University, ul. Ulyanovskaya 1, 198504, Petrodvorets, Saint Petersburg, Russia
}

\abstract{%
  The NA61/SHINE experiment aims to discover the critical point of strongly interacting matter and study the properties of the onset of deconfinement. For these goals a scan of the two dimensional phase diagram (T-$\mu_B$) is being performed at the SPS by measurements of hadron production in proton-proton, proton-nucleus and nucleus-nucleus interactions as a function of collision energy and system size.

In this contribution intriguing results on the energy dependence of hadron spectra and yields in inelastic p+p and centrality selected Be+Be collisions will be presented. In particular, the energy dependence of the signals of deconfinement, the 'horn', 'step' and 'kink', in p+p interactions will be presented and compared with the corresponding results from central Pb+Pb collisions from NA49.
}
\maketitle
\section{Introduction}
\label{intro}
\begin{figure}[h]
\centering
\includegraphics[scale=.35]{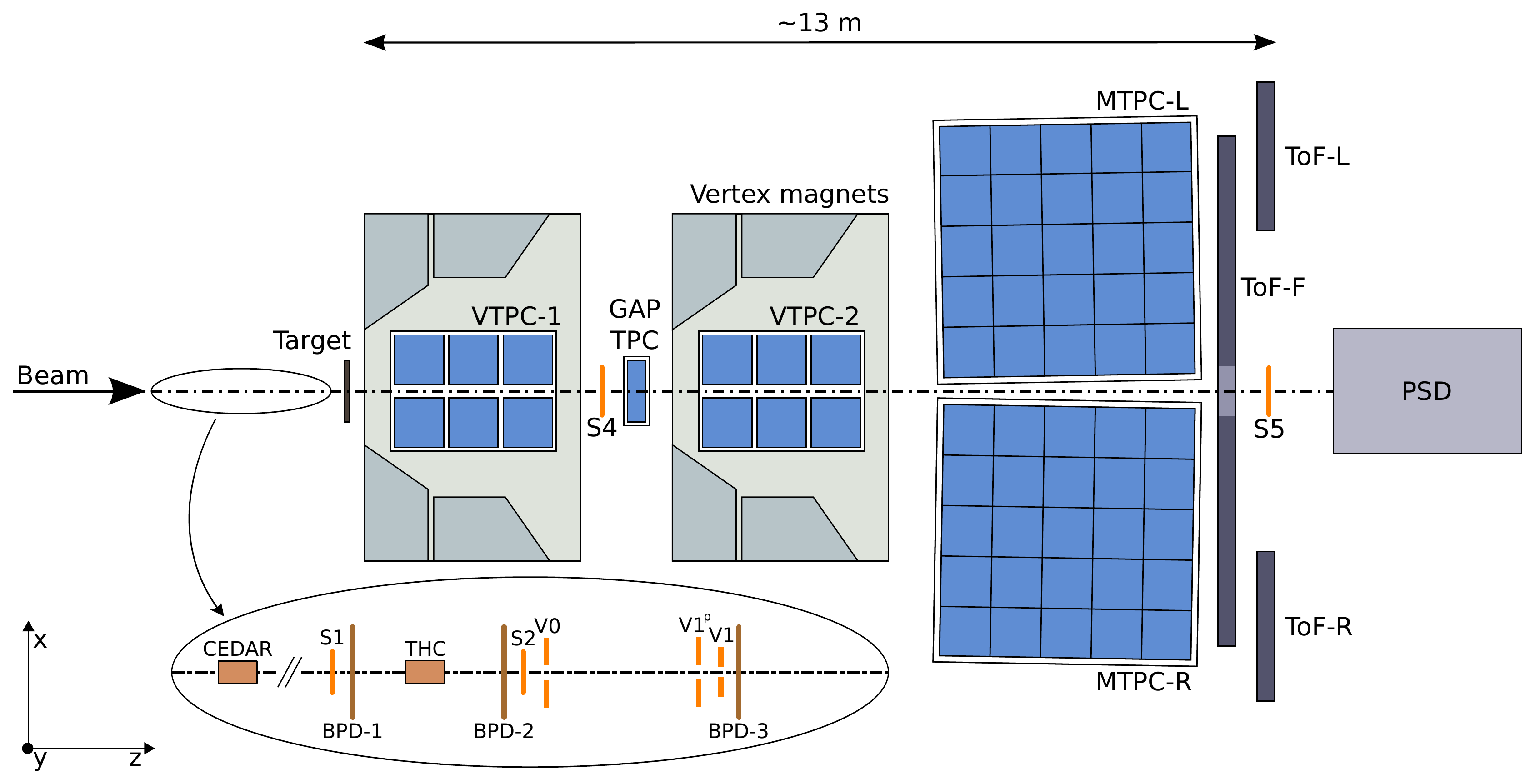}
\caption{Schematic layout of the NA61/SHINE experiment at the CERN SPS
(horizontal cut in the beam plane, not to scale).}
\label{fig1} 
\end{figure}

NA61/SHINE~\cite{NA61} is a fixed target experiment at the Super Proton Synchrotron (SPS) of the European Organization for Nuclear Research (CERN). The layout of the NA61/SHINE detector is sketched in fig.~\ref{fig1}.
It consists of a large acceptance hadron spectrometer with
excellent capabilities in charged particle momentum measurements and
identification by a set of five Time Projection Chambers as well as
Time-of-Flight detectors. The high resolution modular forward calorimeter,
the Projectile Spectator Detector, measures energy flow around the beam
direction, which in nucleus-nucleus reactions is primarily a measure of
the number of spectator (non-interacted) nucleons and thus
related to the centrality of the collision.
Both primary and secondary beams are available to the experiment, allowing data taking with projectile sizes ranging from proton to lead, as well as with pions and kaons. An array of beam detectors identifies beam particles, secondary
hadrons and ions as well as primary ions, and measures precisely
their trajectories.

\begin{figure}[h]
\centering
\includegraphics[width=7cm,clip]{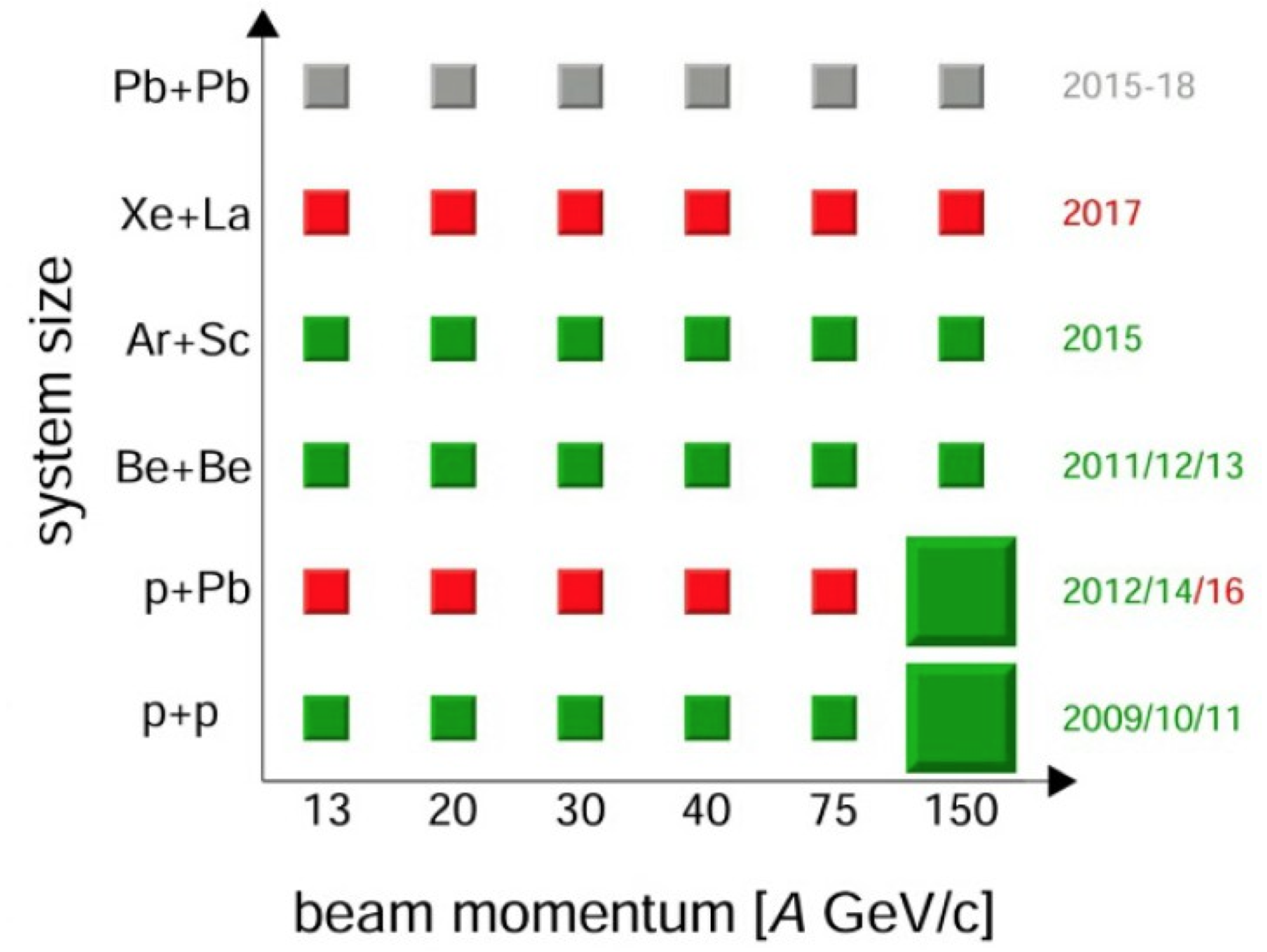}
\caption{Data taking status of the strong interaction programme of NA61/SHINE.}
\label{datatak}       
\end{figure}

The main goal of the strong interaction programme of the experiment is to discover the Critical Point (CP)~\cite{Fodor:2004nz}  of strongly interacting matter
and study the properties of the onset of deconfinement (OD)~\cite{Gazdzicki:1998vd, Alt:2007aa}. To achieve this goal a two-dimensional phase diagram scan - energy versus system size - is being performed by NA61/SHINE. It includes measurements of hadron production in collisions of protons and various nuclei (p+p, Be+Be, Ar+Sc, Xe+La) at a range of beam momenta (13{\it A} - 158{\it A} GeV/c). Figure~\ref{datatak} shows for which systems and energies data has already been collected (green), is
scheduled for recording (red) or is planned (gray). Besides studying strong interactions, the experiment also performs precise hadron production reference measurements for neutrino (Fermilab and T2K) and cosmic-ray (KASCADE and Pierre Auger Observatory) physics.
\section{Analysis of identified hadron spectra}
\label{analysis}
\begin{figure}[h]
\centering
\sidecaption
\includegraphics[width=7cm,clip]{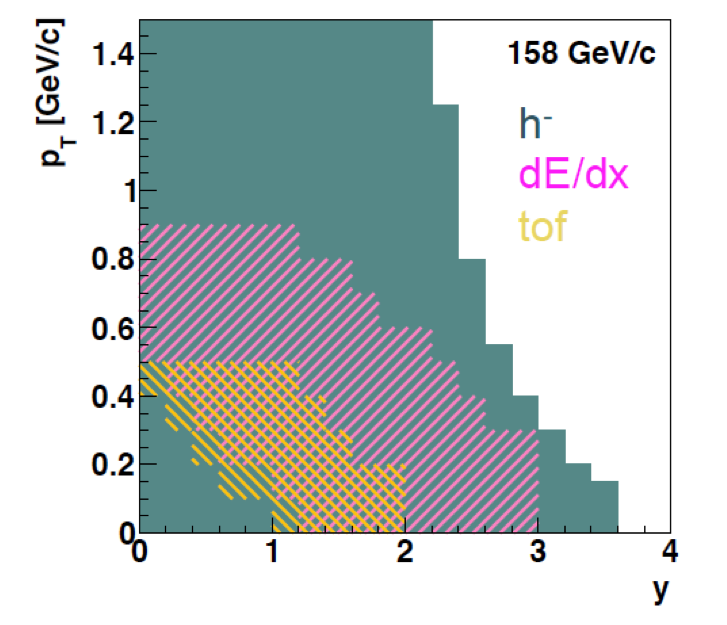}
\caption{Acceptance for various data analysis methods for $\pi^{-}$ from  inelastic p+p interactions at 158 GeV/c. Blue, solid area: h$^-$; magenta shading: $dE/dx$; yellow shading: time-of-flight.}
\label{accept}       
\end{figure}

The identification of particles was performed for inelastic p+p interactions and centrality selected Be+Be collisions using such techniques as the h$^-$ method~\cite{Abgrall:2013qoa}, $dE/dx$ and $tof-dE/dx$ measurements.  The h$^-$ method is based on the fact that the majority of negatively charged hadrons produced in inelastic interactions at SPS energies are pions. The remaining small contamination ($\approx 10\%$) of other hadrons ($K^{-}$ and $\overline{p}$) is subtracted using models of particle production (e.g. EPOS1.99~\cite{Pierog:2009zt}). The $dE/dx$ method is based on the particle energy loss in the TPC gas. It can be applied for $\pi^{+}$, $\pi^{-}$, $K^{+}$, $K^{-}$, $p$ and $\overline{p}$ with total momentum ($p_{tot}$) in the momentum range $4 - 100$ GeV/c and with transverse momentum ($p_{T}$) in the momentum range $0 - 2$ GeV/c. The time-of-flight measurement together with the particle path length and momentum measured by the TPCs allow to calculate its mass $m$.  Combined $tof-dE/dx$ analysis provides excellent separation of different hadron species close to mid-rapidity. It is performed  for $\pi^{+}$, $\pi^{-}$, $K^{+}$, $K^{-}$, $p$ and $\overline{p}$ with $p_{tot}$ in the momentum range $2 - 10$ $GeV/c$ and with $p_{T}$ in the momentum range $0 - 2$ GeV/c. The acceptances in rapidity versus transverse momentum phase space for these methods are shown in fig.~\ref{accept} for p+p interactions at 158~GeV/c. 

The results presented in this paper are corrected for detector inefficiencies, feed-down from weak decays and secondary interactions, contribution from non-target interactions, as well as trigger and event selection biases. The detailed simulation of the NA61/SHINE detector response was based on GEANT3.
\section{p+p results}
\label{p+p}
This section presents inclusive spectra of identified hadrons produced in inelastic p+p interactions
at 20, 31, 40, 80, 158 GeV/c and observables, that are usually treated as signalsof the onset of deconfinement~\cite{Pulawski:2015tka}.

\begin{figure}[h]
\centering
\includegraphics[width=7cm,trim={0 1cm 0 5cm},clip]{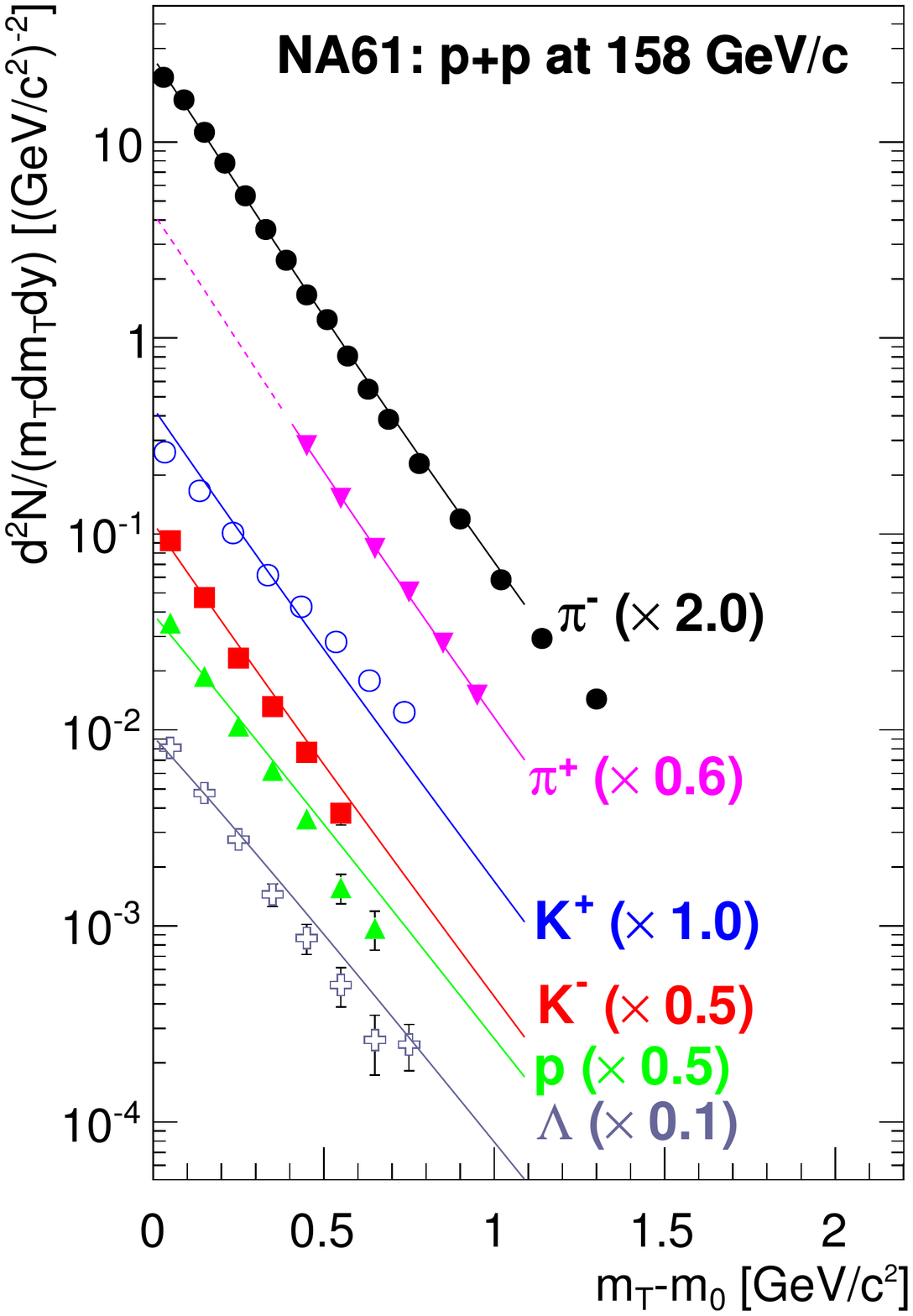}
\includegraphics[width=7cm,trim={0 1cm 0 5cm},clip]{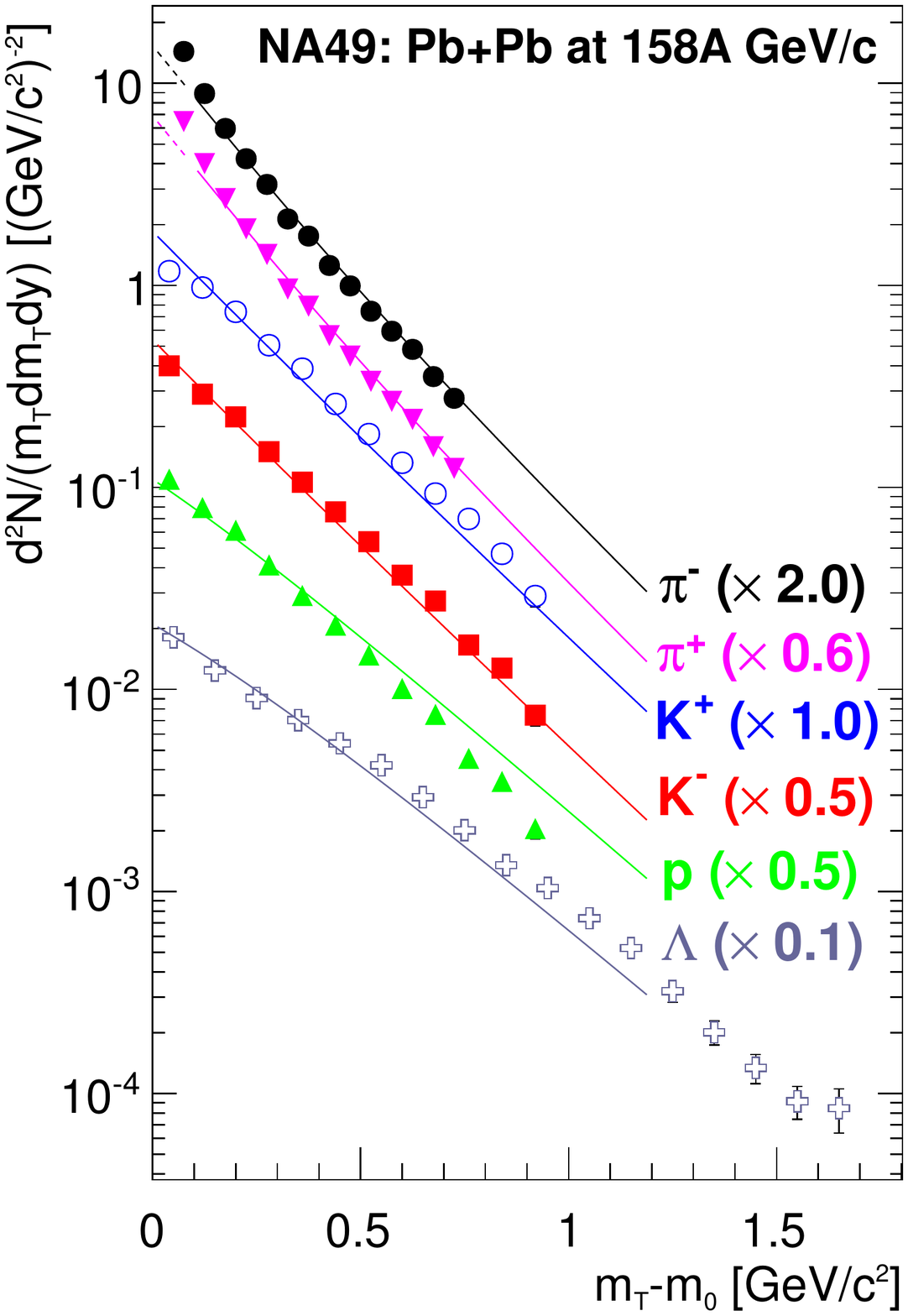}
\caption{Transverse mass spectra at mid-rapidity. Left: NA61/SHINE inelastic p+p interactions at 158~GeV/c; right: NA49 central Pb+Pb collisions at 158{\it A}~GeV/c. Lines show fits based on the blast wave model.}
\label{hagedorn}       
\end{figure}

Figure~\ref{hagedorn} (left) shows spectra of transverse mass of negatively and positively charged $\pi$, $K$, $p$ and $\Lambda$~\cite{Aduszkiewicz:2015dmr} produced in inelastic p+p interactions at mid-rapidity. Corresponding NA49 measurements~\cite{Alt:2007aa,Alt:2006dk,Afanasiev:2002mx} for central Pb+Pb collisions are shown in fig.~\ref{hagedorn} (right). The data was fitted using a blast wave model parametrisation~\cite{Schnedermann:1993ws} $\frac{dN_i}{m_Tdm_Tdy}=A_im_TK_1\left(\frac{m_T\cosh\rho}{T}\right)I_0\left(\frac{p_T\sinh\rho}{T}\right)$, where the parameter $\rho$ is related to the transverse flow velocity $\beta_T$ by $\rho=\tanh^{-1}\beta_T$.
One finds that $\beta_T$ is significantly smaller in p+p than Pb+Pb collisions.
While the spectra are approximately exponential in p+p reactions with inverse slope parameter of the order 150 MeV as expected from the Hagedorn statistical bootstrap model~\cite{Hagedorn:1972}, this exponential dependence is possibly modified in Pb+Pb interactions by the transverse flow.

The analysis of particle spectra is an important step towards the study of the system size dependence of signals of the onset of deconfinement observed in central Pb+Pb collisions ('kink', 'horn' and 'step')~\cite{Alt:2007aa}. These signatures were predicted in the Statistical Model of the Early Stage (SMES)~\cite{Gazdzicki:2010iv}, which suggests that the phase transition line is crossed by heavy ion collisions between the top AGS energy (beam energy 11.7A GeV) and the
top SPS energy (beam energy 158A GeV).

\begin{figure}[h]
\centering
\includegraphics[width=7cm,trim={6.5cm 9.2cm 0 8cm},clip]{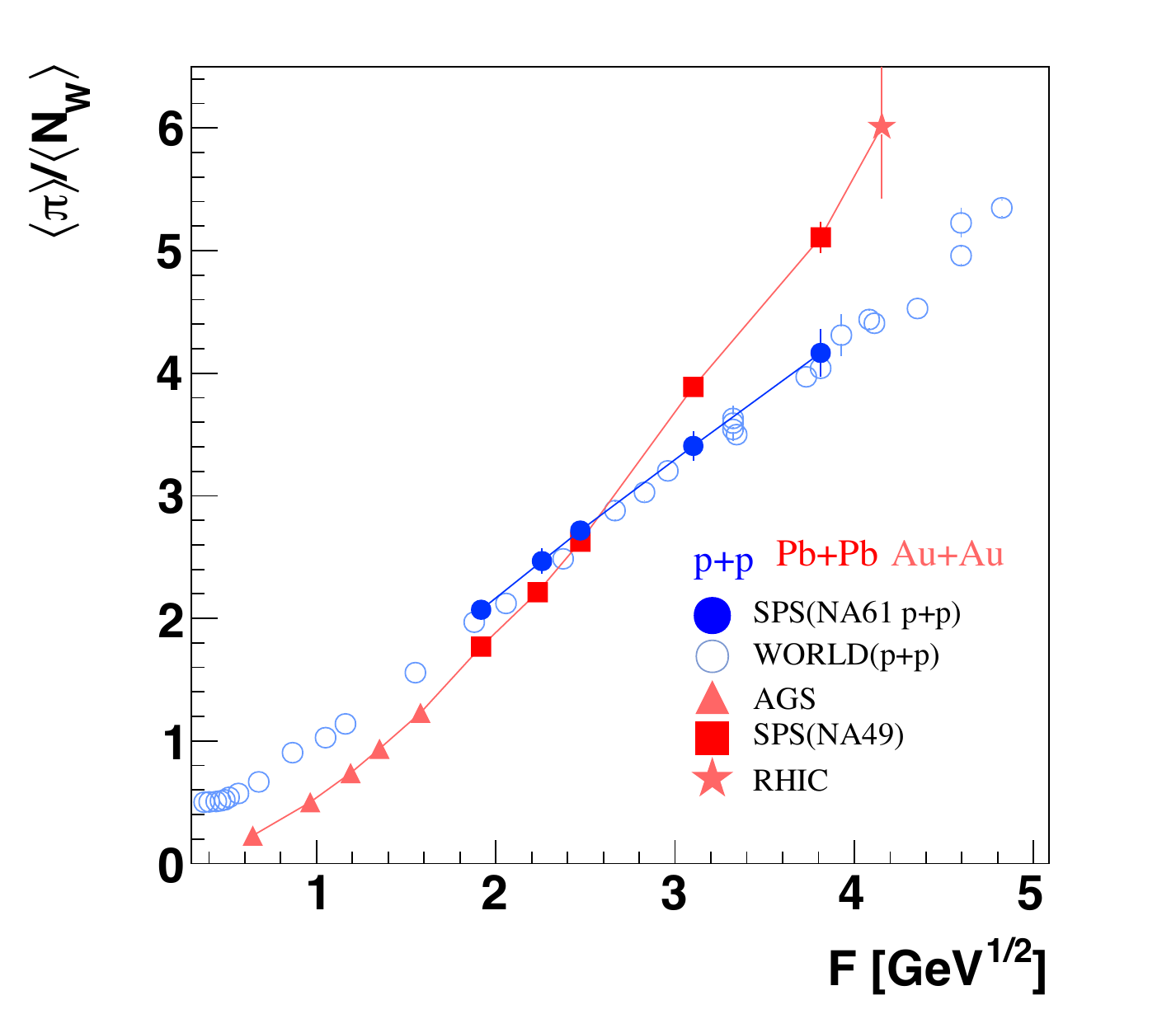}
\caption{The ratio of mean pion multiplicity to the mean number of wounded nucleons for p+p interactions measured by the NA61/SHINE experiment (full blue circles) and other experiments (open blue circles)
and central Au+Au and Pb+Pb collisions plotted as a function of collision energy $F={\left(\frac{{\left(\sqrt{s_{NN}}-2\,m_{N}\right)}^{3}}{\sqrt{s_{NN}}}\right)}^{1/4}$. World data from refs.~\cite{Golokhvastov:2001,Abbas:2013}.}
\label{kink}       
\end{figure}

Figure~\ref{kink} shows that the energy dependence of mean $\pi$ multiplicity increases slower in p+p than in Pb+Pb collisions ('kink').
Hence, the two dependences cross each other at around 40{\it A}~GeV/c, slightly violating the prediction of the Wounded Nucleon Model~\cite{Bialas:1976ed} $\langle\pi\rangle_{(AA)}/\langle N_{W}\rangle = \langle\pi\rangle_{pp}/2$,
while the SMES model predicts that the creation of the QGP leads to the increase of entropy and, consequently, to the rise of the slope for the plotted dependence. The way of obtaining the mean $\pi$ multiplicity in inelastic p+p interactions is described in detail in ref.~\cite{Abgrall:2013qoa}.

\begin{figure}[h]
\centering
\includegraphics[width=7cm,trim={6.5cm 9.5cm 0 8cm},clip]{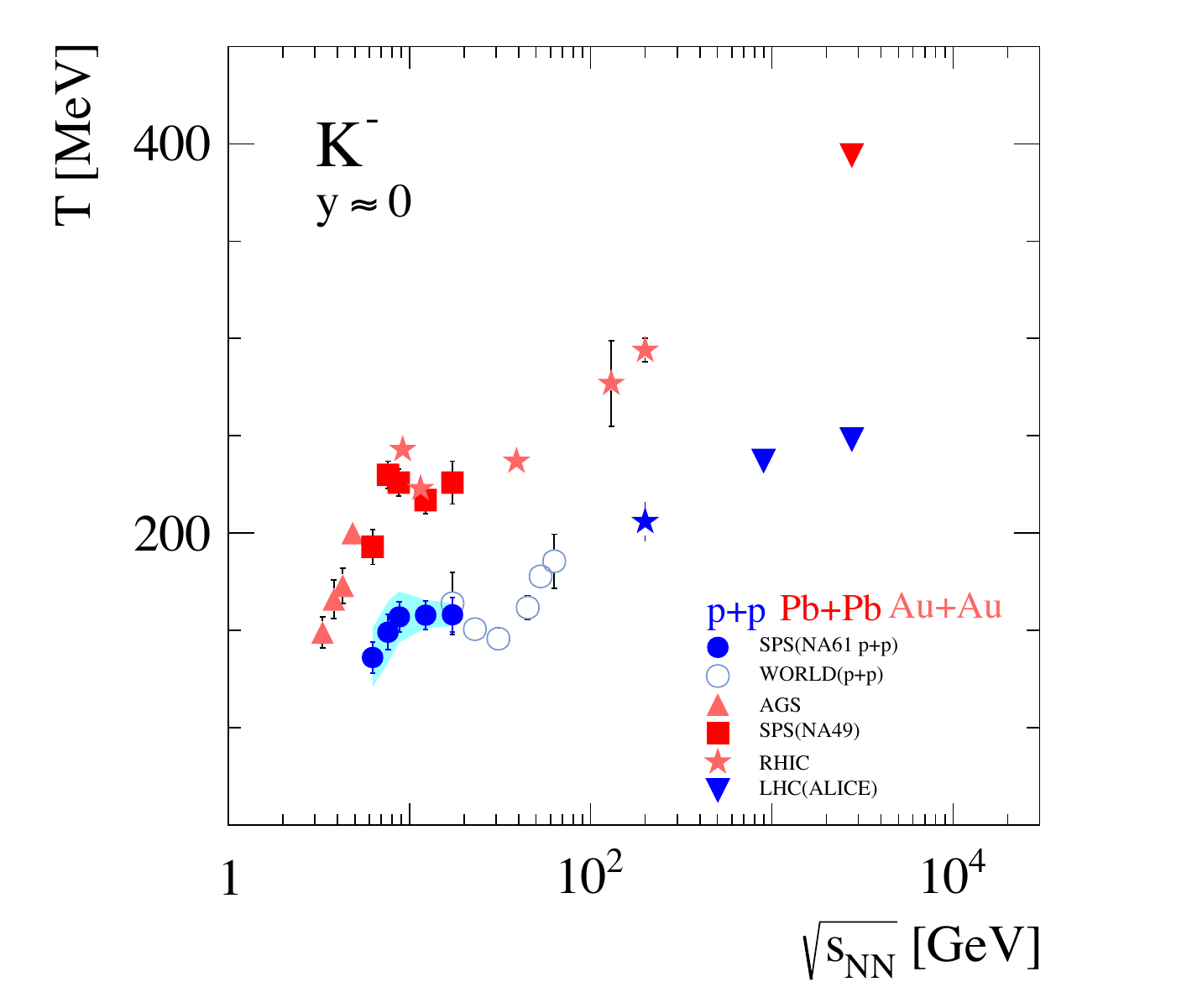}
\includegraphics[width=7cm,trim={6.5cm 9.5cm 0 8cm},clip]{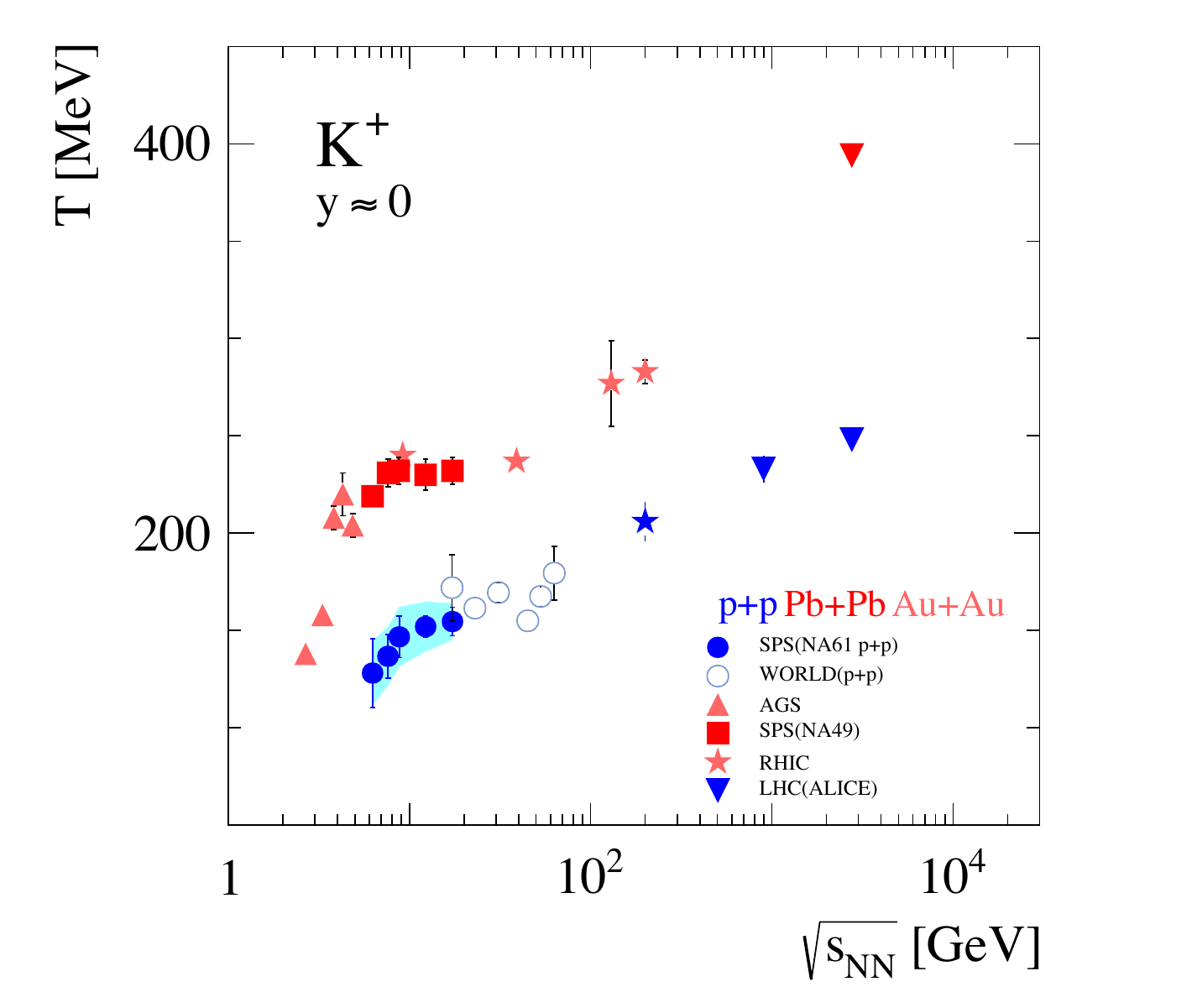}
\caption{Energy dependence of the inverse slope parameter T of transverse mass spectra of K$^-$ and K$^+$ in inelastic p+p interactions measured by the NA61/SHINE experiment (full blue circles) and other experiments (open blue circles, blue star and blue triangles) and central Au+Au and Pb+Pb interactions. World data from refs.~\cite{Kliemant:2003sa,Abelev:2008ab,Aamodt:2011zj,Abelev:2014laa}}
\label{step} 
\end{figure}

The new measurements of NA61/SHINE significantly improve the world data for the inverse-slope parameter $T$ of $m_T$ spectra of kaons.
The $m_T$ spectra were fitted to the exponential function $\frac{d^2n}{dp_Tdy}=\frac{Sp_T}{T^2+m_KT}\exp\left(-\frac{\sqrt{p^2_T+m^2_K}-m_K}{T}\right)$, where $S$ is the yield integral and $m_K$ is the K mass. Figure~\ref{step} shows the energy dependence of $T$ for K$^+$ and K$^-$. Surprisingly the NA61/SHINE results from inelastic p+p collisions exhibit rapid changes like observed in central Pb+Pb interactions and reproduce the 'step'-like behaviour.

\begin{figure}[h]
\centering
\includegraphics[width=7cm,trim={6.5cm 9.5cm 0 8cm},clip]{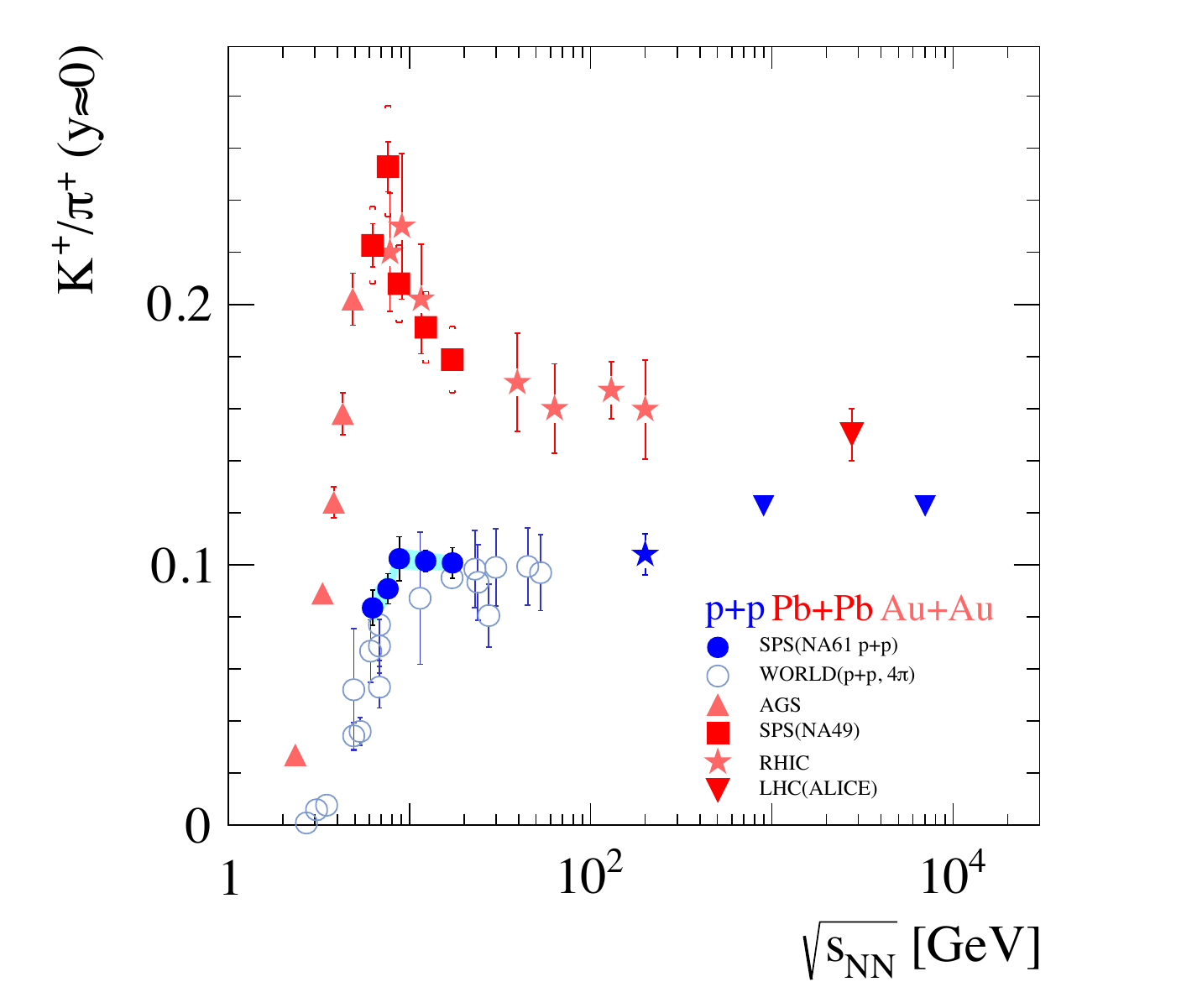}
\includegraphics[width=7cm,trim={6.5cm 9.5cm 0 8cm},clip]{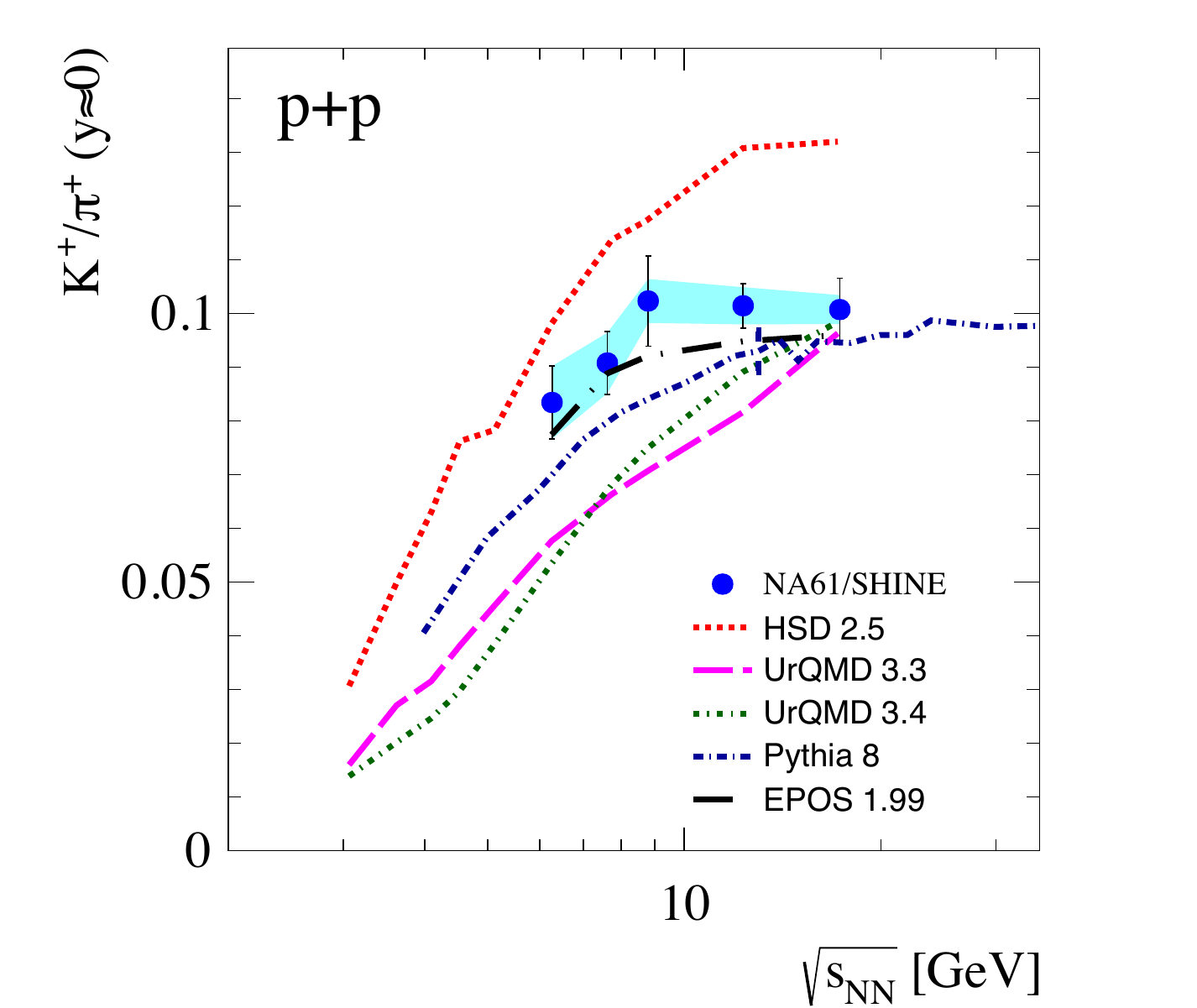}
\caption{Energy dependence of the K$^+$/${\pi}^+$ ratio in inelastic p+p interactions interactions measured by the NA61/SHINE experiment (full blue circles) and other experiments (open blue circles, blue star and blue triangles) compared with central Pb+Pb and Au+Au interactions (left) and theoretical model predictions (right). World data from refs.~\cite{Aamodt:2011zj,Gazdzicki:1995zs,Gazdzicki:1996pk,Arsene:2005mr,Abelev:2012wca}, models~\cite{Pierog:2009zt,Ehehalt:1996,Bass:1998,Bleicher:1999,Sjostrand:2015}.}
\label{horn}       
\end{figure}

The energy dependence of the K$^+$/${\pi}^+$ ratio at mid-rapidity for inelastic p+p interactions and central Pb+Pb/Au+Au collisions is presented in fig.~\ref{horn} (left). Again surprisingly, the data suggests that even inelastic p+p interactions exhibit a step-like structure in the energy dependence of the K$^+$/${\pi}^+$ ratio, which looks like a precursor of the 'horn' structure observed for Pb+Pb collisions in the same energy domain. This finding can be qualitatively explained by the modified SMES model when a phase transition is assumed in the p+p interactions and exact strangeness conservation is taken into account~\cite{Poberezhnyuk:2015}. Furthermore, the measured ratios were compared to theoretical models, as shown in fig.~\ref{horn} (right). It is evident that none of the models quantitatively describes the data, with EPOS1.99 being the most consistent.

\begin{figure}[h]
\centering
\includegraphics[width=7cm,trim={6.5cm 9.5cm 0 8cm},clip]{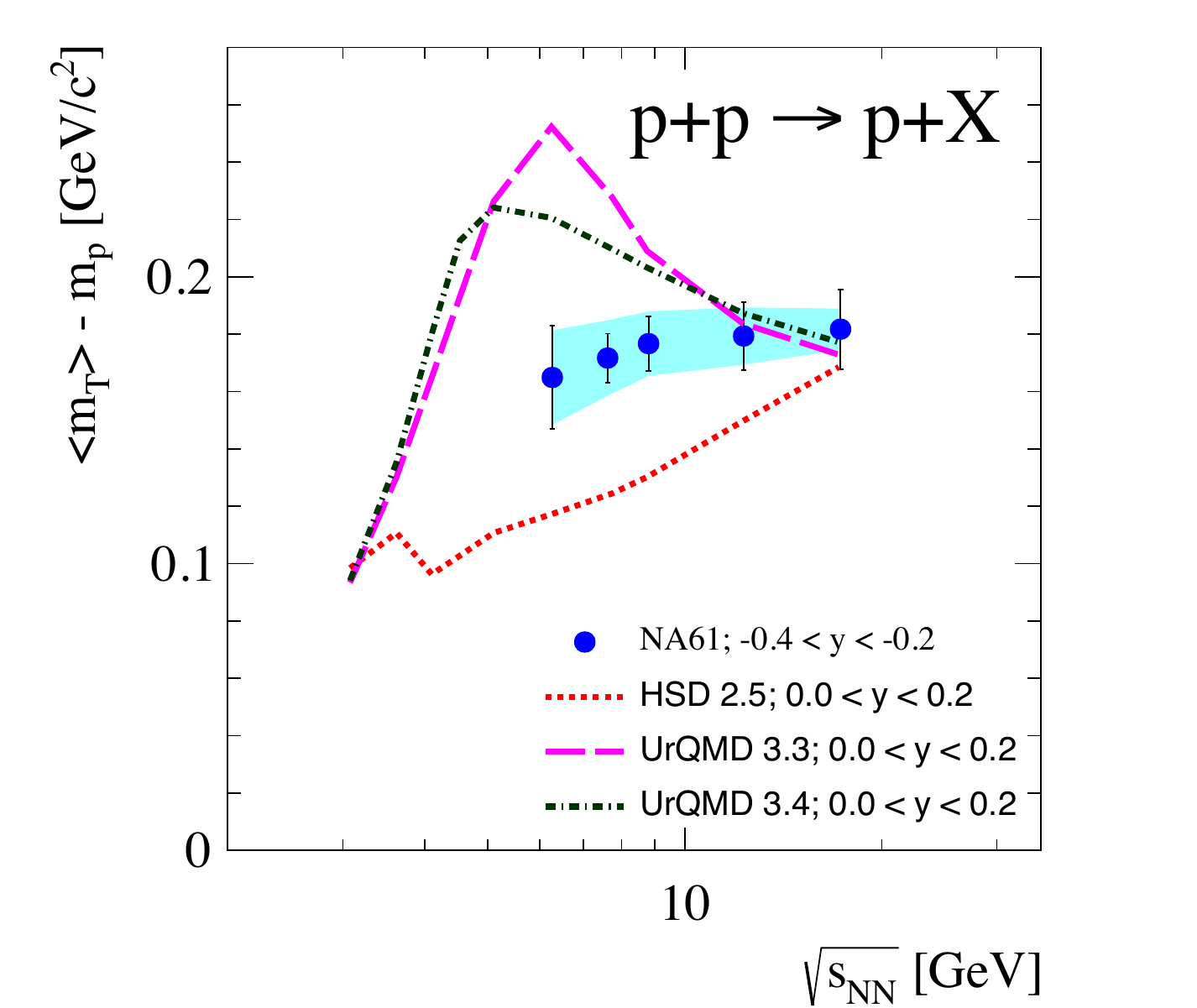}
\caption{Comparison of energy dependence of $\langle m_T\rangle-m_p$ in inelastic p+p interactions with predictions of models~\cite{Vovchenko:2014}.}
\label{protons}       
\end{figure}

Proton transverse momentum spectra and mean proton transverse mass $\langle m_{T}\rangle$, around mid-rapidity, in inelastic p+p interactions were also studied by the NA61/SHINE collaboration. The results are compared with theoretical models in fig.~\ref{protons}. Mean $m_T$ of protons was calculated from the following parametrization of data: $\frac{d^2n}{dp_Tdy}=\frac{Sp_T}{T^2+m_pT}\exp\left(-\frac{\sqrt{p^2_T+m^2_p}-m_p}{T}\right)$. It exhibits a slow increase with the collision energy in contradiction to the UrQMD or HSD models predictions, that do not reproduce this dependence neither qualitatively nor quantitatively.
\section{Be+Be results}
\label{Be+Be}
Previously, the inelastic cross section for $^7$Be+$^9$Be was only measured at 1.45{\it A}~GeV/c~\cite{Tanihata:1986kh}.
The new measurements of NA61/SHINE allowed to extend this to 13{\it A}, 20{\it A} and 30{\it A}~GeV/c~\cite{weimer2013}.
These measurements were performed using a round scintillator counter S4 placed on the beam-line behind the target, that measures the square of the charge of a particle passing through it.
Inelastic interactions were selected by requiring a signal below that expected for an intact beam nucleus.
Data was also taken with the target removed to be able to subtract the background caused by beam interactions with detector material.

\begin{figure}[h]
\centering
\includegraphics[width=14cm,trim={4cm 12cm 4cm 12cm},clip]{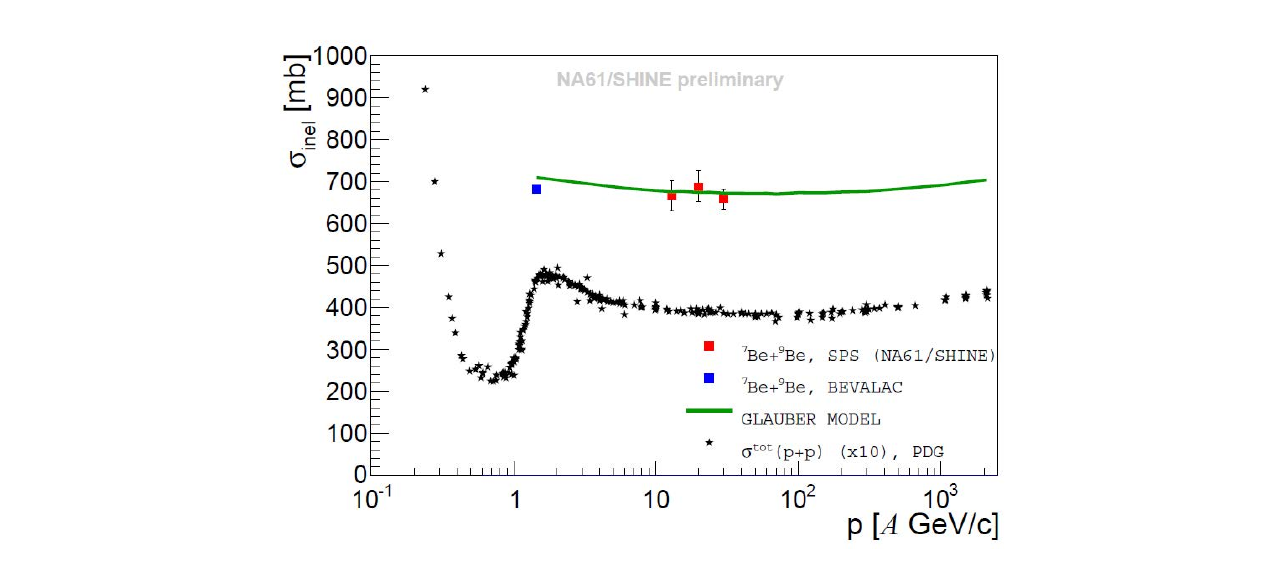}
\caption{Inelastic cross-section of $^7$Be+$^9$Be interactions as a function of beam momentum in comparison with the total cross-section of p+p interactions.}
\label{cross}       
\end{figure}

The interaction probability in the target is given by $P_{int}=\frac{P_I-P_R}{1-P_R}$, where $P_I$ and $P_R$ are the interaction probabilities when the target is inserted and removed, respectively.
Using $P_{int}$, the cross section can be calculated from $\sigma_{inel}=\frac{1}{\rho L_{eff} N_A/A}P_{int}$, where  $L_{eff}=\lambda_{abs}(1-e^{-L/\lambda_{abs}})$ and $\lambda_{abs}=\frac{A}{\rho N_A\sigma_{inel}}$.
$N_A$, $\rho$, $L$ and $A$ are the Avogadro constant, the target density, length and atomic number, respectively.
Figure~\ref{cross} shows that the new measurements are in good agreement with the previous measurement as well as the Glauber model~\cite{Broniowski:2007nz} prediction.

\begin{figure}[h]
\centering
\includegraphics[width=4.5cm,trim={6.5cm 9cm 0 8cm},clip]{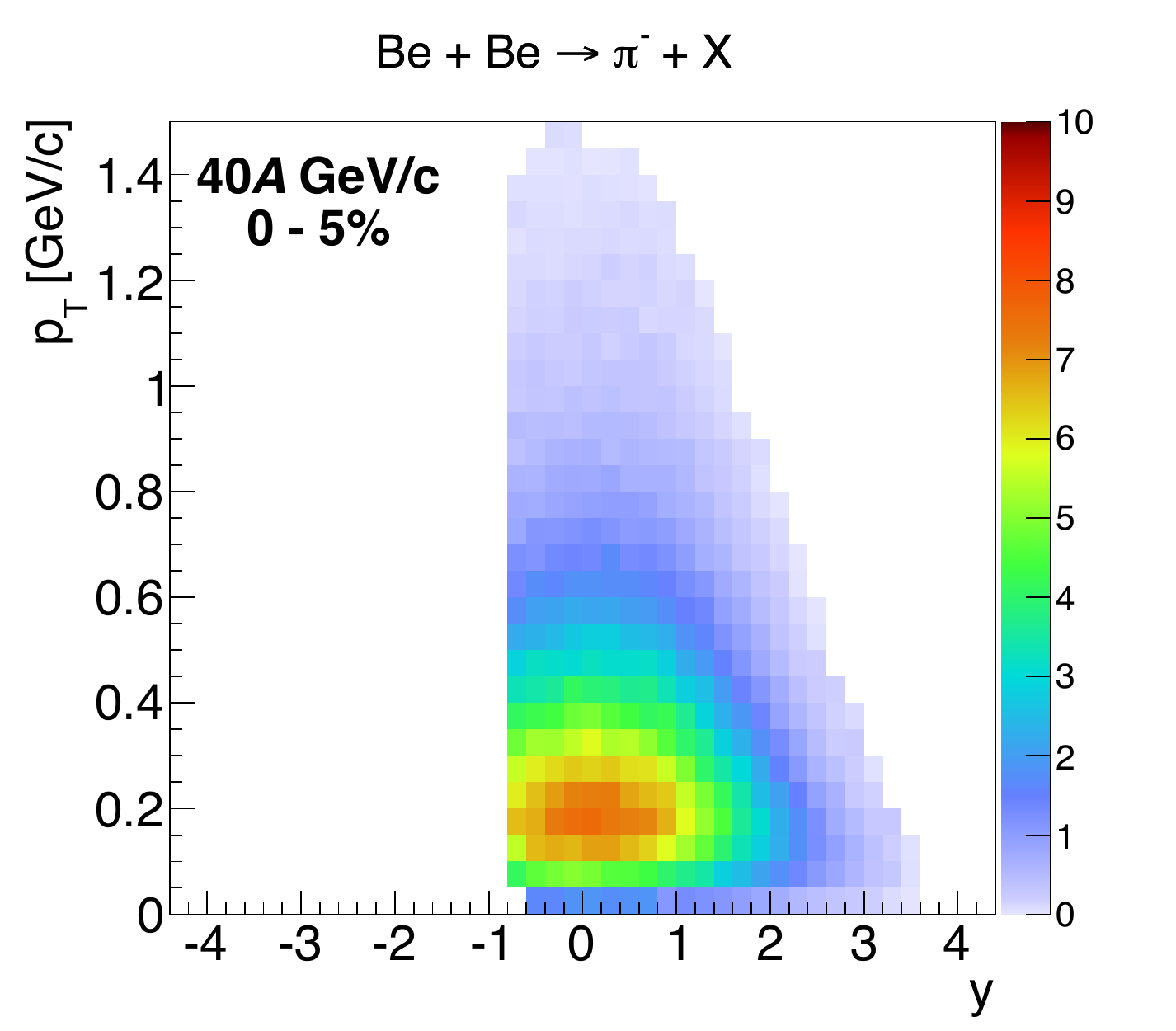}
\includegraphics[width=4.5cm,trim={6.5cm 9cm 0 8cm},clip]{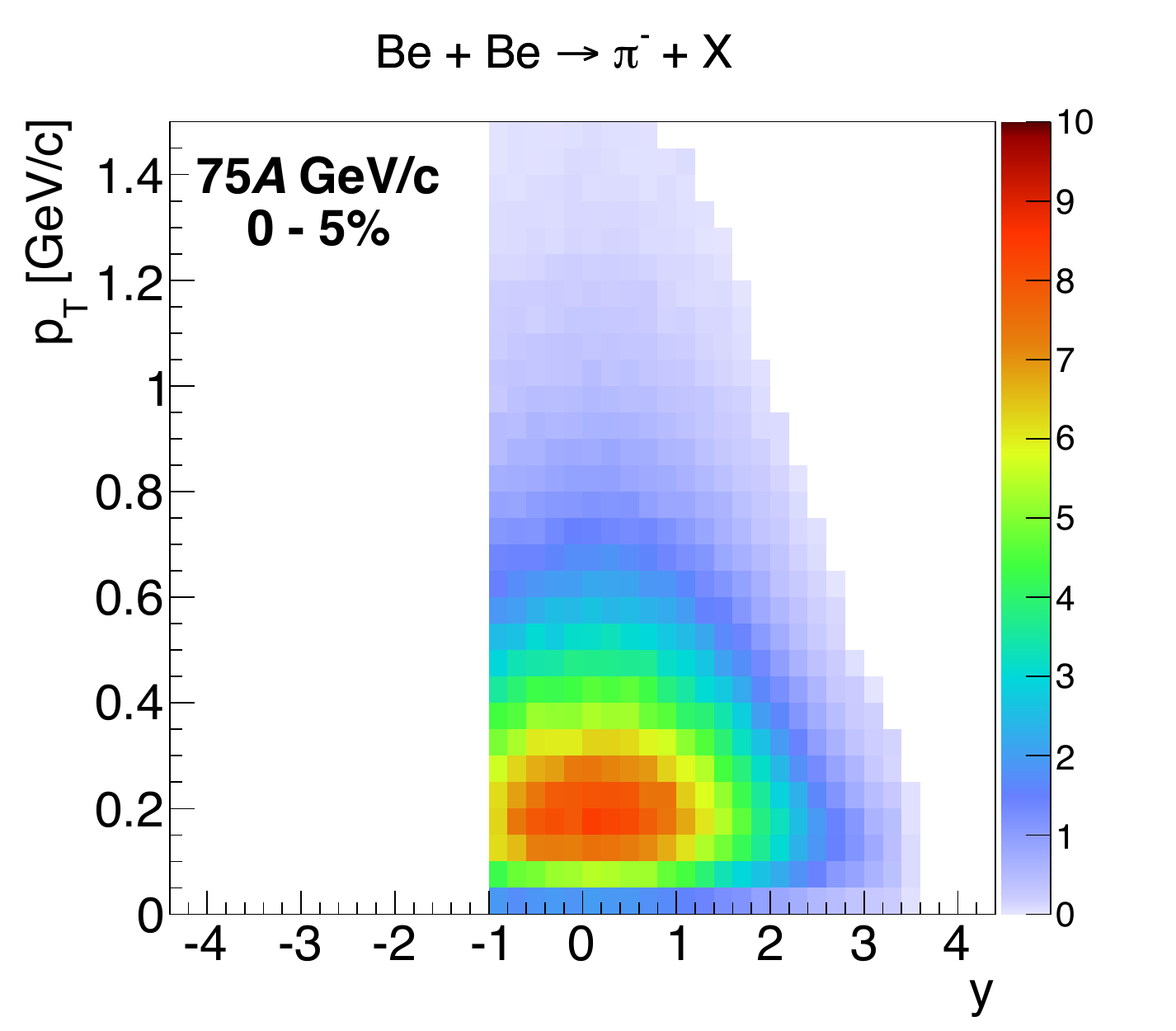}
\includegraphics[width=4.5cm,trim={6.5cm 9cm 0 8cm},clip]{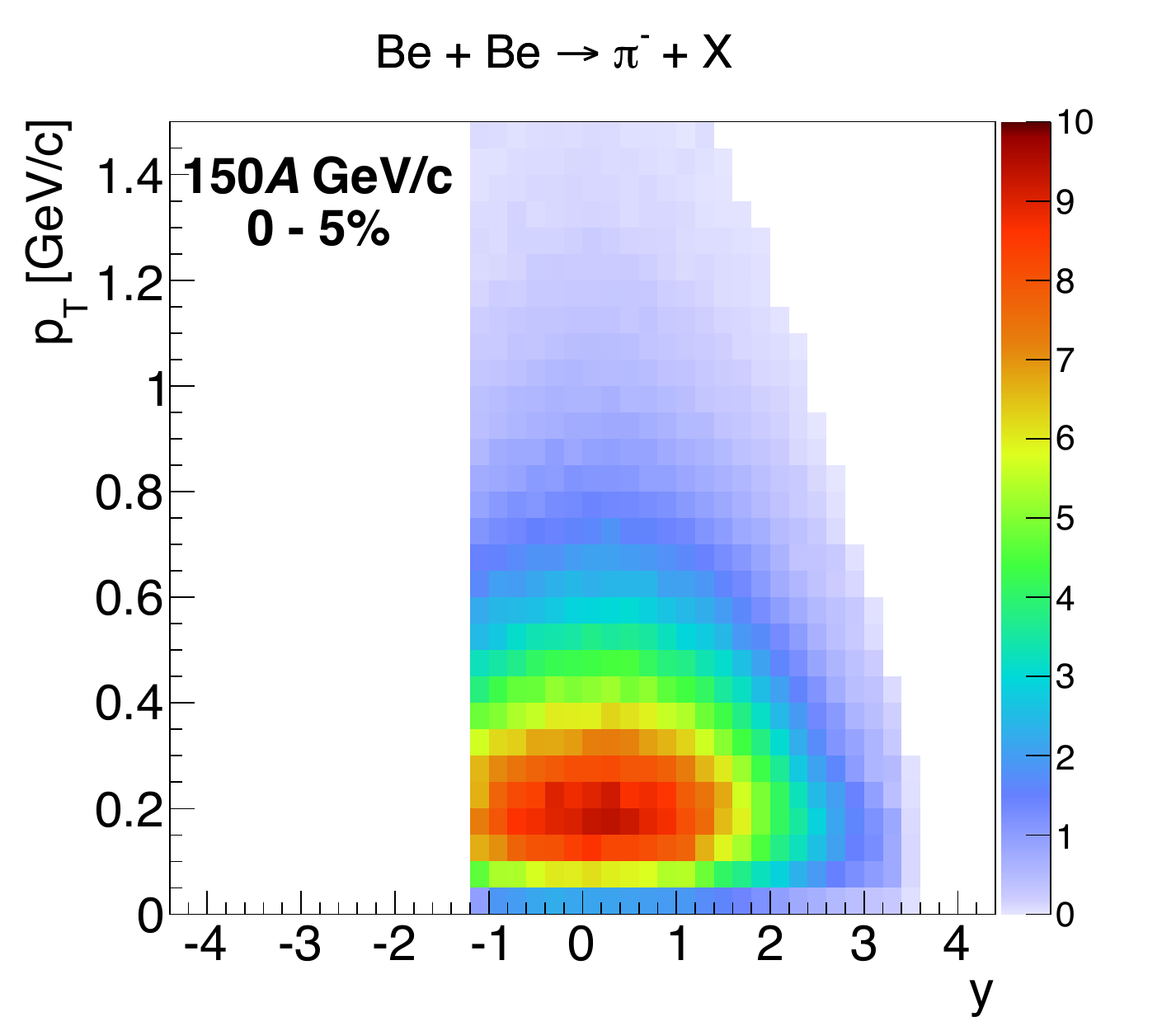}\\
\includegraphics[width=4.5cm,trim={6.5cm 9cm 0 8cm},clip]{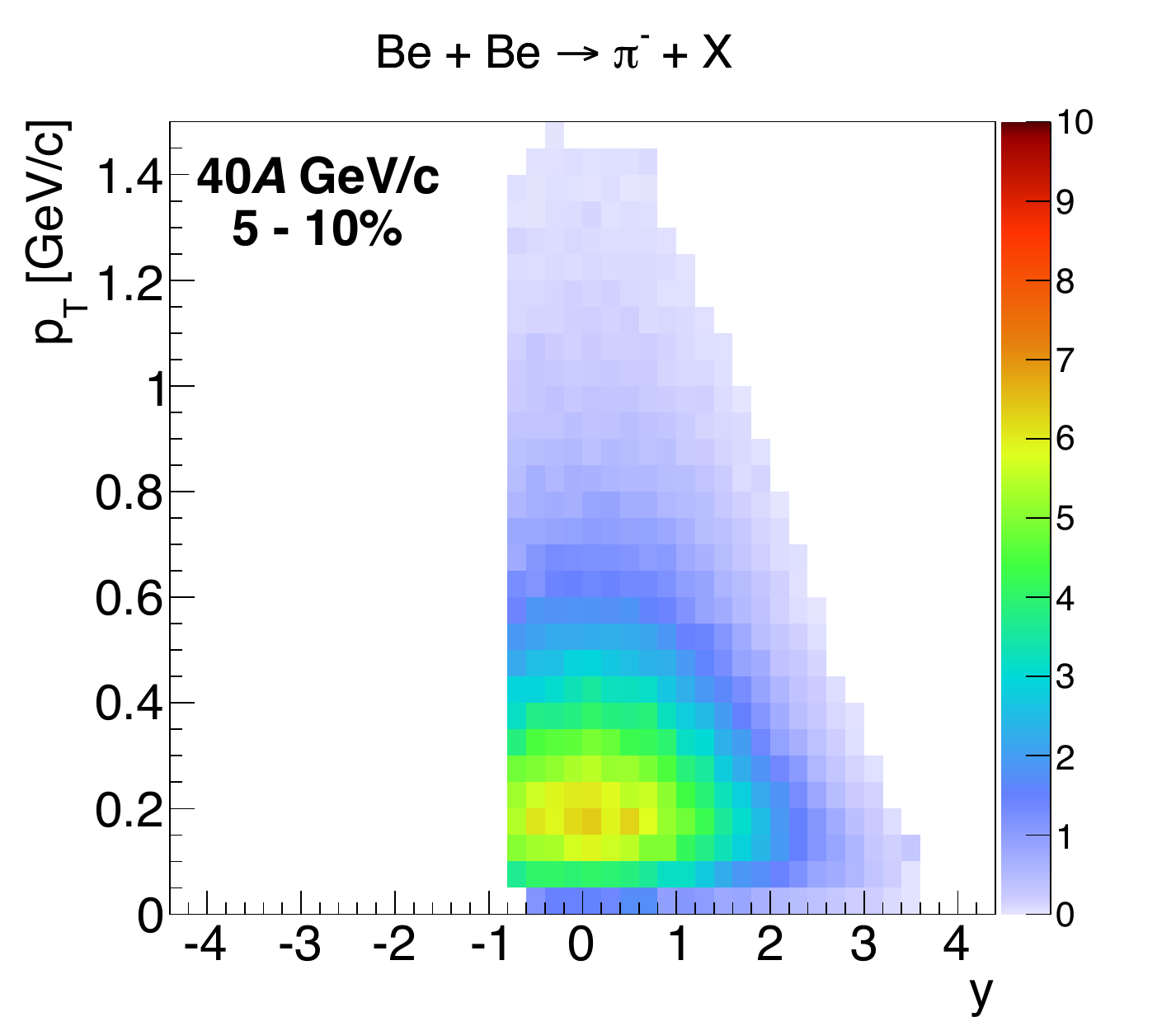}
\includegraphics[width=4.5cm,trim={6.5cm 9cm 0 8cm},clip]{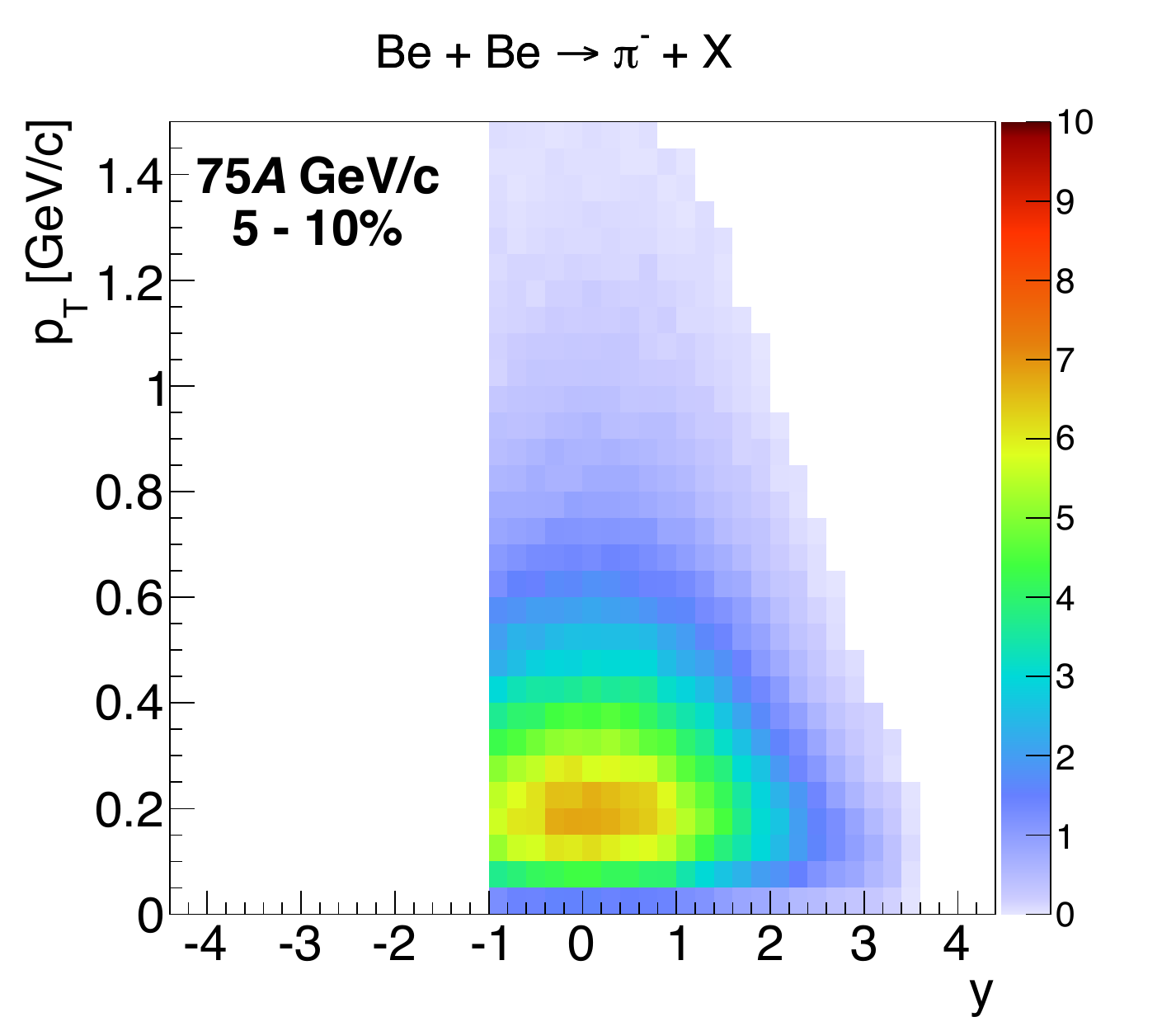}
\includegraphics[width=4.5cm,trim={6.5cm 9cm 0 8cm},clip]{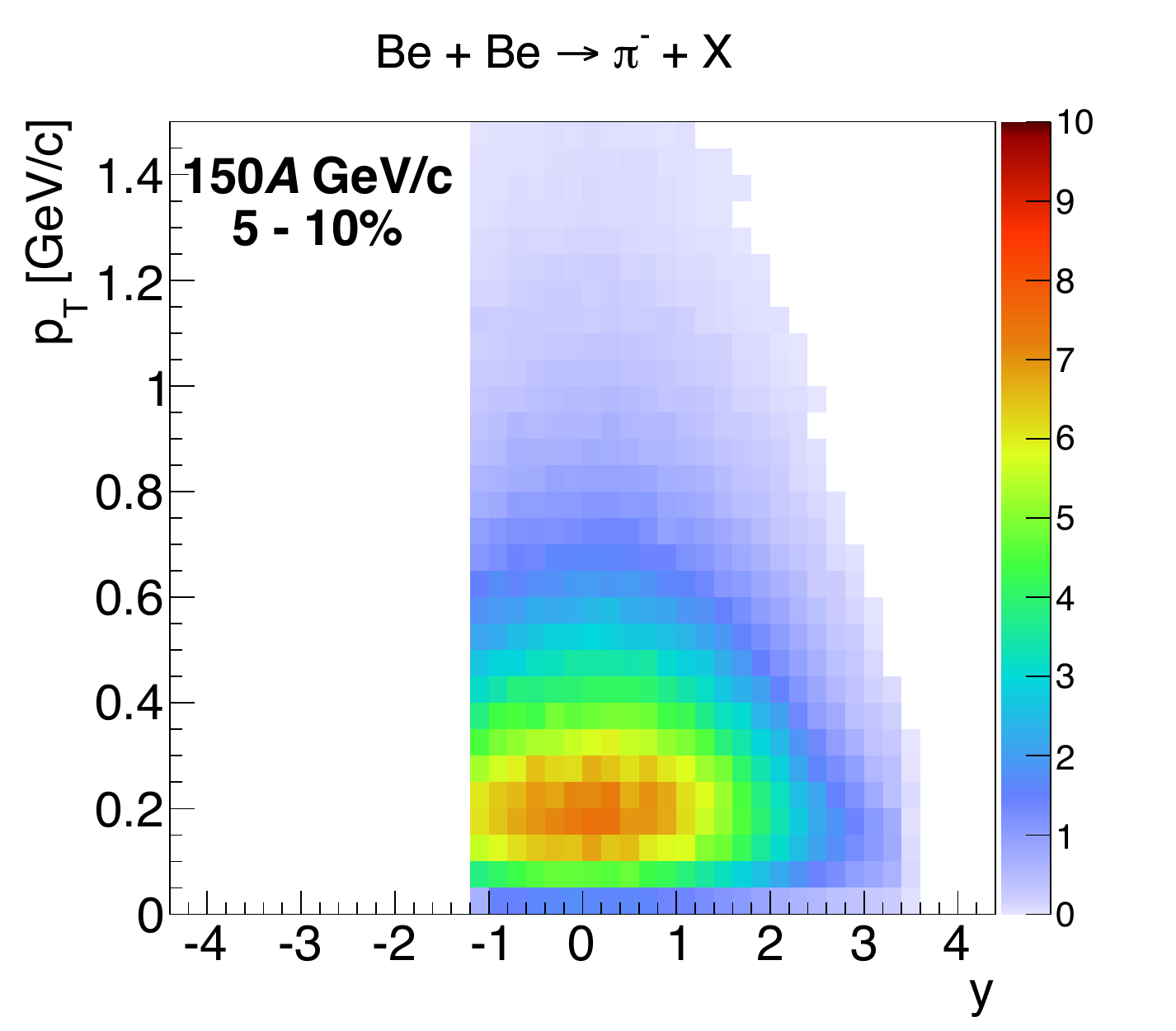}
\caption{Double differential spectra of negatively charged pions in rapidity and transverse momentum for $^7$Be+$^9$Be collisions at 40{\it A}, 75{\it A} and 150{\it A} GeV/c for 0-5$\%$ and 5-10$\%$ centrality classes.}
\label{spectra}       
\end{figure}

Preliminary results on spectra were obtained for $^7$Be+$^9$Be interactions at beam momenta of 20{\it A}, 30{\it A}, 40{\it A}, 75{\it A} and 150{\it A}~GeV/c in the centrality classes 0--5\%, 5--10\%, 10--15\% and 15--20\%~\cite{Kaptur:2014}. The double differential spectra of $\pi^{-}$ in rapidity and transverse momentum for $^7$Be+$^9$Be collisions at three beam momenta and in two centrality classes are presented in fig.~\ref{spectra}. The centrality was derived from the energy deposited in the forward calorimeter, the Projectile Spectator Detector.

\begin{figure}[h]
\centering
\includegraphics[width=7cm,trim={0 0 0 8cm},clip]{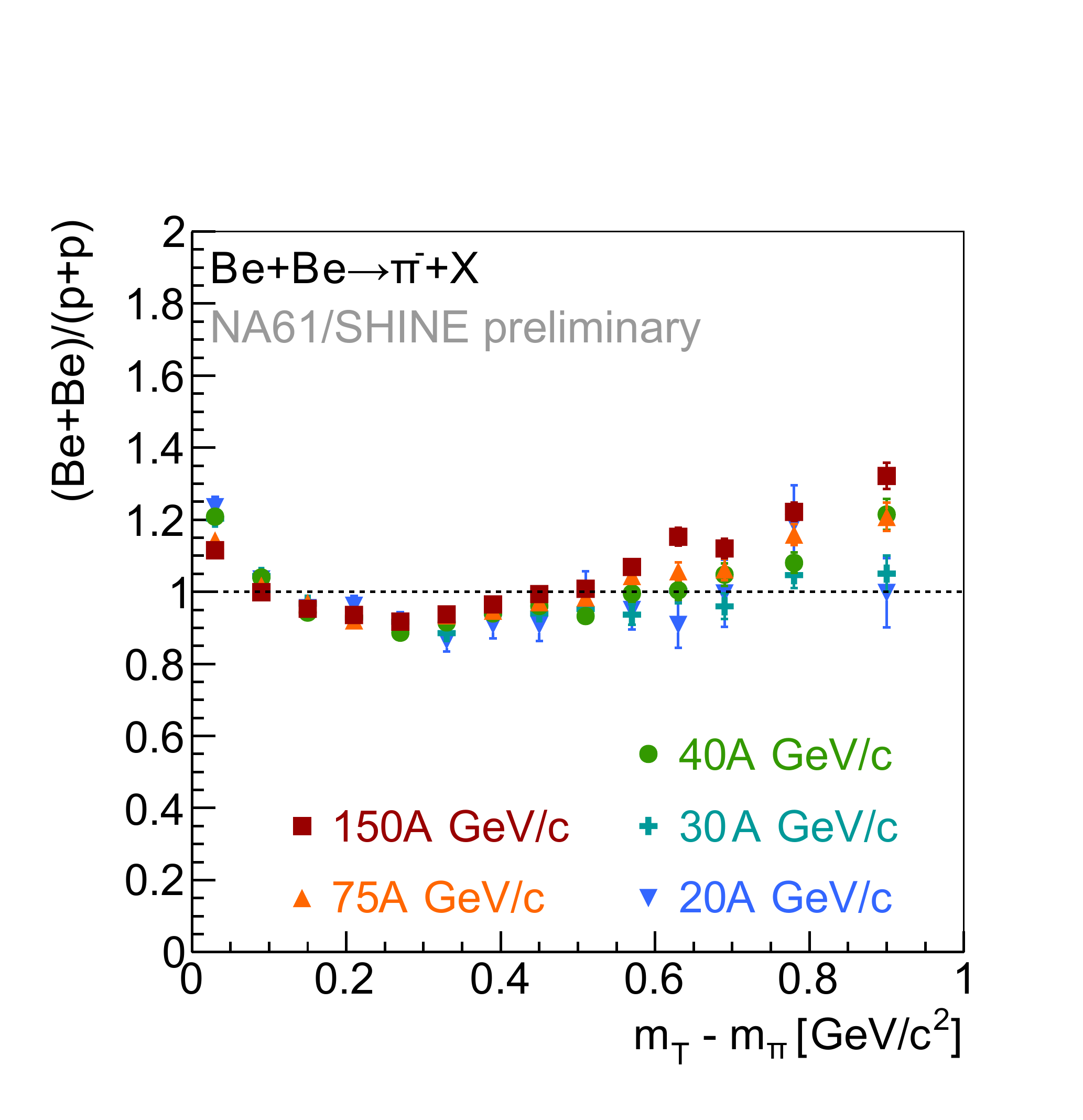}
\includegraphics[width=7cm,trim={6.5cm 9cm 0 8cm},clip]{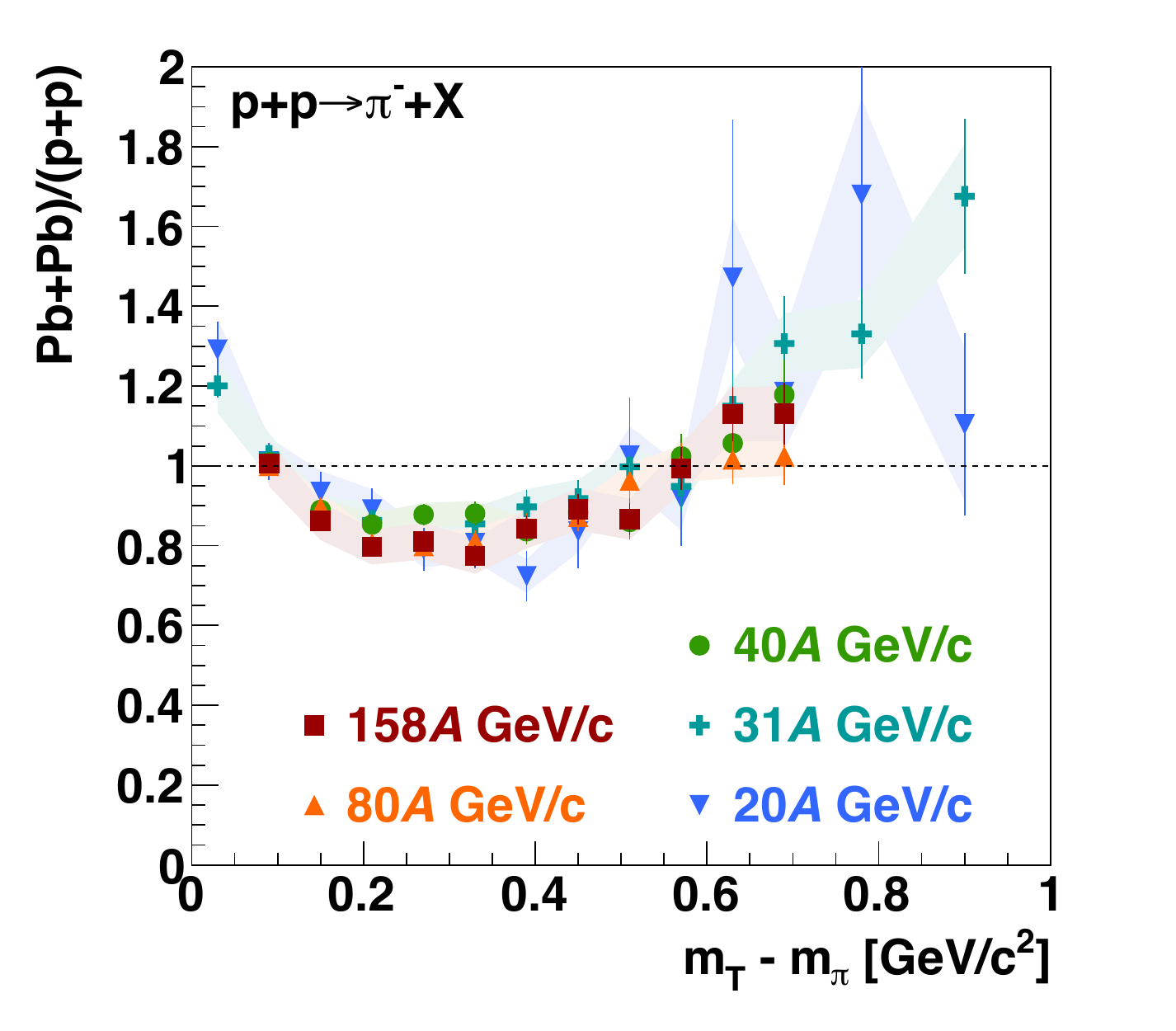}
\caption{Ratio of normalised transverse mass spectra of $\pi^{-}$ mesons for different system sizes. Left: ($^7$Be+$^9$Be)/(p+p) (15$\%$ of the most central events); right: (Pb+Pb)/(p+p) (5$\%$ of the most central events for 150{\it A} GeV/c and 7.5$\%$ of the most central events for other energies~\cite{Abgrall:2013qoa})}
\label{RAA}       
\end{figure}

To compare the shape of transverse mass spectra of negatively charged pions for $^7$Be+$^9$Be and Pb+Pb reactions the ratio of normalized transverse mass spectra for $^7$Be+$^9$Be/p+p and Pb+Pb/p+p was calculated. Figure~\ref{RAA} shows qualitatively similar behaviour for both $^7$Be+$^9$Be (15$\%$ of the most central events) and Pb+Pb reactions (5$\%$ of the most central events for 150{\it A} GeV/c and 7.5$\%$ of the most central events for other energies).

The high-$m_T$ regions of both $^7$Be+$^9$Be and Pb+Pb exhibit an increase of the ratio, while for the intermediate regions a decrease is seen. This effect is stronger for central Pb+Pb collisions. Usually, this is attributed to collective flow. Also, for $^7$Be+$^9$Be reactions the increase of the ratio at high values of $m_T$ appears to be larger at higher beam momenta. This may suggest an increase of the magnitude of collective effects in $^7$Be+$^9$Be collisions with increasing beam momentum. The low-$m_T$ regions of both $^7$Be+$^9$Be and Pb+Pb reactions exhibit an increase of the ratio. Possible explanations include isospin asymmetry of p+p data or electromagnetic effects.
\section{Conclusions}
\label{conc}
The ongoing NA61/SHINE scan program covers a wide range of beam energies and collision system sizes. The analysis of the first results from inelastic p+p and $^7$Be+$^9$Be interactions at CERN SPS energies showed many interesting effects.
In particular, the p+p data exhibited step-like structures in the energy region where the onset of deconfinement was found in central Pb+Pb collisions.
Moreover, it was observed that many of the measurements could not be explained well by theoretical models.
For $^7$Be+$^9$Be reactions new results on the inelastic cross section were obtained at several energies. Transverse mass and rapidity spectra were measured in the SPS energy range for several centrality classes. An indication of a transverse flow effect was found at the highest beam momenta.
\begin{acknowledgement}
This work was supported by
the Federal Agency of Education of the Ministry of Education and Science of the
Russian Federation, SPbSU research grant 11.38.193.2014,
the Hungarian Scientific Research Fund (grants OTKA 68506 and 71989),
the J\'anos Bolyai Research Scholarship of
the Hungarian Academy of Sciences,
the Polish Ministry of Science and Higher Education (grants 667\slash N-CERN\slash2010\slash0, NN\,202\,48\,4339 and NN\,202\,23\,1837),
the Polish National Center for Science (grants~2011\slash03\slash N\slash ST2\slash03691, 2012\slash04\slash M\slash ST2\slash00816 and 
2013\slash11\slash N\slash ST2\slash03879),
the Foundation for Polish Science --- MPD program, co-financed by the European Union within the European Regional Development Fund,
the Russian Academy of Science and the Russian Foundation for Basic Research (grants 08-02-00018, 09-02-00664 and 12-02-91503-CERN),
the Ministry of Education, Culture, Sports, Science and Tech\-no\-lo\-gy, Japan, Grant-in-Aid for Sci\-en\-ti\-fic Research (grants 18071005, 19034011, 19740162, 20740160 and 20039012),
the German Research Foundation (grant GA\,1480/2-2),
the EU-funded Marie Curie Outgoing Fellowship,
Grant PIOF-GA-2013-624803,
the Bulgarian Nuclear Regulatory Agency and the Joint Institute for
Nuclear Research, Dubna (bilateral contract No. 4418-1-15\slash 17),
Ministry of Education and Science of the Republic of Serbia (grant OI171002),
Swiss Nationalfonds Foundation (grant 200020\-117913/1)
and ETH Research Grant TH-01\,07-3.
\end{acknowledgement}

%
%
%

\end{document}